\documentclass[12pt]{iopart}

\usepackage[british]{babel}
\usepackage{iopams}
\usepackage{graphicx}

\def\be{\begin{equation}}
\def\ee{\end{equation}}
\def\bea{\begin{eqnarray}}
\def\eea{\end{eqnarray}}
\def\ba{\begin{array}}
\def\ea{\end{array}}

\newcommand{\eqref}[1]{equation (\ref{#1})}
\newcommand{\figref}[1]{figure \ref{#1}}
\newcommand{\tabref}[1]{table \ref{#1}}
\newcommand{\secref}[1]{section \ref{#1}}

\def\Mpc{{\rm Mpc}}

\def\muka{$\mu$K\,arcmin}

\newcommand{\Om}{\Omega_{\rm m}}
\newcommand{\Od}{\Omega_{\rm d}}

\newcommand{\Ode}{\Omega_{\rm d}^e}
\newcommand{\Omh}{\Omega_{\rm m}^0 h^2}
\newcommand{\om}{\omega_{\rm m}}
\newcommand{\Obh}{\Omega_{\rm b}^0 h^2}
\newcommand{\ob}{\omega_{\rm b}}

\def\Dplus{D_{+}}
\newcommand{\V}[1]{\mbox{\boldmath $#1$}}
\def\zmean{z_{\rm mean}}
\def\fsky{f_{\rm sky}}
\newcommand{\dd}{\mathrm{d}}

\begin{document}

\title[Constraints on early dark energy from CMB lensing and weak lensing tomography]
{Constraints on early dark energy from CMB lensing and weak lensing tomography}

\author[Lukas Hollenstein, Domenico Sapone, Robert Crittenden and Bj{\"o}rn Malte Sch\"afer]
{Lukas Hollenstein$^1$, Domenico Sapone$^2$, Robert Crittenden$^1$ and Bj{\"o}rn Malte Sch\"afer$^{3,4,1}$}

\address{$^1$ Institute of Cosmology \& Gravitation, University of Portsmouth, Dennis Sciama Building, Burnaby Road, Portsmouth PO1 3FX, United Kingdom}

\address{$^2$ D\'epartement de Physique Th\'eorique, Universit\'e de Gen\`eve, 24, Quai Ernest-Ansermet, CH--1211 Gen\`eve 4, Switzerland}

\address{$^3$Astronomisches Recheninstitut, Zentrum f{\"u}r Astronomie, Universit{\"a}t Heidelberg, M{\"o}nchhofstra{\ss}e 12, 69120 Heidelberg, Germany}

\address{$^4$ Institut d'Astrophysique Spatiale, Universit{\'e} de Paris XI, b{\^a}timent 120-121, Centre universitaire d'Orsay, 91405 Orsay CEDEX, France}

\ead{
\mailto{lukas.hollenstein@port.ac.uk},
\mailto{domenico.sapone@unige.ch},
\mailto{robert.crittenden@port.ac.uk},
\mailto{spirou@ita.uni-heidelberg.de}
}

\begin{abstract}
Dark energy can be studied by its influence on the expansion of the Universe as well as on the growth history of the large-scale structure. In this paper, we follow the growth of the cosmic density field in early dark energy cosmologies by combining observations of the primary CMB temperature and polarisation power spectra at high redshift, of the CMB lensing deflection field at intermediate redshift and of weak cosmic shear at low redshifts for constraining the allowed amount of early dark energy. We present these forecasts using the Fisher-matrix formalism and consider the combination of Planck-data with the weak lensing survey of Euclid. We find that combining these data sets gives powerful constraints on early dark energy and is able to break degeneracies in the parameter set inherent to the various observational channels. The derived statistical $1\sigma$-bound on the early dark energy density parameter is $\sigma(\Ode)=0.0022$ which suggests that early dark energy models can be well examined in our approach. In addition, we derive the dark energy figure of merit for the considered dark energy parameterisation and comment on the applicability of the growth index to early dark energy cosmologies.
\end{abstract}



\section{Introduction} \label{sec:intro}

Cosmological models with a dark energy fluid with an equation of state parameter $w$ close to $-1$ are favoured by combining recent Cosmic Microwave Background (CMB), supernova and Baryon Acoustic Oscillations (BAO) data, suggesting that the Hubble-expansion accelerates in the current cosmic epoch.  However, the behaviour of this dark energy fluid at early times is still an open question; if its equation of state remains close to $-1$ at early times, such as for a cosmological constant, its density evolves slowly and was neglible relative to the matter and radiation fluids.  In this case, the dark energy is unlikely to play a role in the physics of the early universe. 

However, in dynamic models of dark energy, such as a scalar field, this may be different and dark energy could have a non-negligible influence on earlier stages of the growth history. It has been shown that scalar field models exist with global attractor solutions which sub-dominantly ``track'' the dominant component of the cosmological fluid \cite{Wetterich:1987fk,Ratra:1987rm,Zlatev:1998tr,Steinhardt:1999nw}.   In such models, the fraction of dark energy fluid to the critical density, $\Ode$, was more or less constant at early times.  If  $\Ode$ was sufficiently large, its presence could have had an observable impact on the probes of the early universe, such as nucleosynthesis, CMB physics and the growth rate of structure.  These models are collectively known as early dark energy (EDE) models.

EDE models may help to resolve some apparent discrepencies in the amplitude of density fluctuations, usually parameterised by $\sigma_8$, the r.m.s.~of the density field on scales of $8\ h^{-1}\Mpc$.   If the excess in the CMB power spectrum at high multipoles measured with the Cosmic Background Imager (CBI) \cite{Readhead:2004gy} is explained by the Sunyaev-Zel'dovich (SZ) effect of unresolved clusters of galaxies one needs a value of $\sigma_8$ close to one, which is significantly higher than currently estimated from combining CMB and lensing data \cite{Bond:2002tp}.  Secondly, the observed strong lensing cross section is predicted by $\Lambda$CDM or traditional dark energy models to be too small by an order of magnitude \cite{Bartelmann:1997xa, Meneghetti:1999au, Wambsganss:2003yr, 2005IAUS..225..193D} and one would be forced to increase $\sigma_8$, again to values close to unity \cite{Fedeli:2008jk}.  Analytical studies of spherical collapse in early dark energy models \cite{Bartelmann:2005fc, Sadeh:2007iz} indicate that these models exhibit decreased threshold redshifts for the formation of objects, which would enhance the SZ-signal, although recent studies question this result \cite{Francis:2008md, Grossi:2008xh}.

By combining the currently measured CMB temperature anisotropy power spectrum, supernova data and large-scale clustering from the Sloan Digital Sky Survey (SDSS), Doran \etal\cite{Doran:2006xp} derive upper bounds on the early dark energy parameter $\Ode$ of $0.04$ ($2\sigma$), in carrying out a simultaneous 8-parameter fit while employing a weak prior on spatial flatness and by imposing a lower bound of $-1$ on the dark energy equation of state parameter $w$. Linder \etal\cite{Linder:2008nq} confirm the relevance of early dark energy on the CMB power spectrum if $\Ode$ is larger than $\sim0.03$, and emphasise that ignoring early dark energy can severly bias the determination of $w$ from BAO.  To improve these constraints, Das \& Spergel \cite{Das:2008am} propose a cross-correlation of the CMB and large-scale structure lensing signal, from which they derive a geometrical quantity consisting of the ratio of angular diameter distances. Although this quantity is well suited to investigate dark energy models, it is rather insensitive to the value of $\Ode$. Xia \& Viel \cite{Xia:2009ys} use the latest CMB, BAO and SNIa with a prior on $H_0$ from the Hubble Space Telescope to constrain the amount of early dark energy at the last scattering surface to be less than $2\%$ and in the structure formation era, around matter radiation equality, to be of the order of $6\%$. Adding Gamma Ray Burst and Ly$\alpha$ forest data can tighten the constraints at the last scattering surface by about an order of magnitude.

In this paper, we consider the combination of an observation of the primary CMB fluctuations, the CMB lensing signal and weak cosmic shear for following the growth history in early dark energy models at high, intermediate and low redshift. Cosmic structure growth on the relevant scales is close to linear, where the theory is able to provide reliable predictions.  The motivation of this investigation was to follow the cosmic growth history by combining high-precision probes at recombination redshifts, at intermediate redshifts of $z\simeq4$ by CMB lensing and at low redshifts around unity from weak cosmic shear. As experimental data for the microwave sky we consider the Planck survey, and compare to the corresponding constraints from the CMBPol satellite concept \cite{Baumann:2008aj, Smith:2008an}.  For the weak lensing power spectra we use the characteristics for a satellite mission like Euclid \cite{Refregier:2003ct}.

While this work was in its final stages, the work of de Putter \etal\cite{dePutter:2009kn} appeared which similarly uses the lensed CMB temperature and polarisation power spectra to constrain early dark energy at high redshifts, but combine this data with current supernova samples rather than future weak lensing data.  They also investigate the degeneracies between the neutrino mass and the amount of early dark energy, because of their opposite influence on the dark matter power spectrum.  Our predictions for the EDE constraints coming from lensed CMB data alone appear comparable, given the differences in our assumptions about the experiments.
 
This article is organised as follows: the parameterisation of early dark energy models used in this paper is introduced in \secref{sec:growth} followed by a discussion of the large-scale structure growth. We introduce the modelling of the three observational channels in \secref{sec:theory}. Constraints are presented in \secref{sec:Results} and a summary in \secref{sec:summary} concludes the article.

\section{Early dark energy and structure growth} \label{sec:growth}

\subsection{Parameterisation of dark energy}

\begin{figure}
\begin{center}
\includegraphics[width=0.65\textwidth]{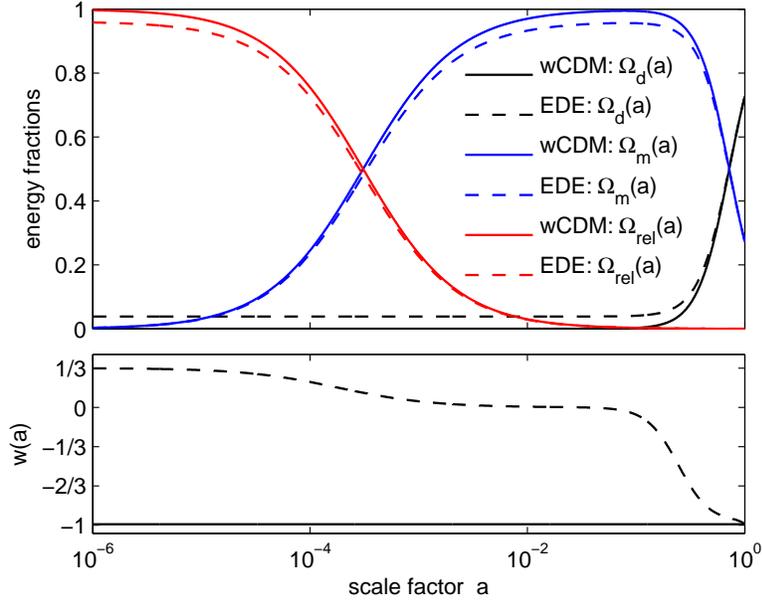}
\caption{The background evolution of the relativistic species (red), baryonic and dark matter (blue), and dark energy (black) are compared in the WMAP+BAO+SN best fit wCDM model (solid) and the EDE parameterisation (dashed) with $\Ode=0.04$. In the lower panel we plot the evolution of the equation of state parameter $w(a)$ corresponding to the two considered models.}
\label{fig:ede_bg}
\end{center}
\end{figure}

For concreteness, we use a  simple phenomenological parameterisation of for a dark energy fluid with an early contribution proposed by Doran \& Robbers \cite{Doran:2006kp}:
\be
\Od(a)\ \equiv\ \frac{\Od^0 - \Ode \left(1- a^{-3 w_0}\right) }{\Od^0 + \Om^0a^{3w_0}}
+\Ode \left (1- a^{-3 w_0}\right) \,.
\label{eq:ede_param}
\ee
Here $\Od^0$ is the fractional energy density today while $\Ode$ represents early contribution. The EDE energy density will sub-dominantly track the evolution of the dominant component. The corresponding equation of state evolves from $w(a\ll 1) = 1/3$ during radiation era to equality and from $w(a>a_eq)=0$ during matter domination to $w(a=1)=w_0$ today, as illustrated by \figref{fig:ede_bg}.   In true scalar field EDE models, the fraction of dark energy is typically proportional to $(1+w_d)$, where $w_d$ is the equation of state of the dominant component;  thus, it is reduced  by $25\%$ as the universe goes from the radiation dominated regime to the matter dominated regime.  Also, the precise transition of the dark energy to its late time behaviour will depend on the details of the scalar potential.  However, while only approximate, this parameterisation should capture most of the essential EDE physics.

Throughout this work we assume a flat cold dark matter model with adiabatic and Gaussian initial conditions. To isolate the EDE effects, we compare these models to those with a dark energy fluid with constant equation of state (wCDM). We assume the following parameters with the WMAP+BAO+SN best fit values from Komatsu \etal\cite{Komatsu:2008hk}:
\bea
& \V{p} = \Big\{ & n_{s}=0.96,\ \tau=0.084,\ \om \equiv\Omh=0.1358,\ \ob\equiv\Obh=0.02267,
\nonumber \\
& &  \Om=0.2732,\ \sigma_{8}=0.8049,\ w_{0}=-0.992,\ \Ode=0.03 \Big\} \,, 
\eea
while the wCDM model is identical but with $\Ode=0$.  Here, $n_s$ is the spectral index of the primordial power spectrum, $\tau$ the optical depth of reionisation, $\om$ and $\ob$ are the physical matter and baryon energy fractions, respectively, $\Om$ is the current fractional energy density of matter, $\sigma_8$ is the normalisation of the matter power spectrum, $w_0$ is the current value of the equation of state of dark energy, and $\Ode$ is the amount of EDE as defined in \eqref{eq:ede_param}. For simplicity, the helium fraction is kept fixed at $Y_{He}=0.24$. From these parameters we derive the Hubble constant $H_0=70.5\,{\rm km /s/Mpc}$, the current dark energy fraction $\Od^0=0.7268$, the amplitude of the primordial power spectrum $\ A_s=3.22\times 10^{-9}$, the reionisation redshift $\ z_{\rm reion}=10.3$, and the age of the Universe $t_0=13.6\,{\rm Gyr}$.

Because the equation of state of dark energy is time dependent we always consider perturbations in the dark energy. For an adiabatic sound speed of $c_s^2=1$ the fluid model with the above EDE parameterisation yields equivalent results as tracking scalar field models. In case of clustering dark energy we find that $c_s^2<1$ introduces a tilt in the matter power spectrum in presence of EDE. The effect of changing $c_s^2$ strongly depends on $\Ode$;  for simplicity, here we only consider the $c_s^2=1$ case.

\subsection{The growth of fluctuations}

An important indicator of the presence of dark energy at any epoch is through its effect on the growth of fluctuations in dark matter.  
At late times, it is well known that the acceleration of the scale factor due to dark energy makes it more difficult for structures to collapse. However, even at early times, when the dark energy density scales identically to the matter density and there is no acceleration, dark energy still suppresses the matter growth because it does not cluster on small scales.

On sub-horizon scales, the linear growth of structure is usually described by the scale-invariant dark matter perturbation normalised today, $\Dplus(a)\equiv \delta_{\rm dm}(a,k) / \delta_{\rm dm}(a=1,k)$. It evolves according to the growth equation
\be
\frac{\dd^2\Dplus}{\dd a^2} +\frac{1}{a}\left[3+\frac{\dd\ln H}{\dd\ln a}\right] \frac{\dd\Dplus}{\dd a}
-\frac{3}{2a^2}\Om(a)\Dplus\ =\ 0 \,.
\label{eq:growth}
\ee
The second term describes the damping of structure growth due to the expansion of the universe.  Early dark energy increases the amount of Hubble damping without contributing to the density perturbation sourcing the growth, so the growth rate is slowed.  In a standard cold dark matter model without dark energy $\Dplus\propto a$, but in the presence of early dark energy, $\Dplus\propto a^{1-3\Ode/5}$ during matter domination \cite{Doran:2001rw,Ferreira:1997hj}.

\begin{figure}
\begin{center}
\includegraphics[width=0.65\textwidth]{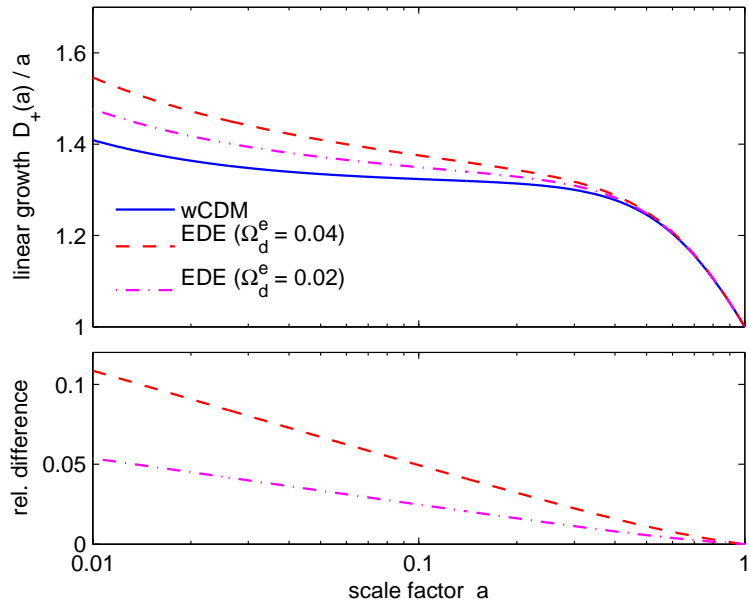}
\caption{Growth of matter perturbations compared for the WMAP+BAO+SN best fit wCDM model (blue, solid) and for the same model with additional EDE with $\Ode=0.04$ (red, dashed) and with $\Ode=0.02$ (magenta, dash-dotted). We observe that  at redshifts $z<5$ ($a>0.25$) the effect of EDE on the growth of perturbations is at the percent level.}
\label{fig:growth_camb}
\end{center}
\end{figure}

This treatment is only approximate, as it does not apply on superhorizon scales and ignores the radiation dominated regime. In practice we use a modified version of the CAMB code\footnote{\texttt{www.camb.info} \cite{Lewis:1999bs}} to evolve the growth numerically. In \figref{fig:growth_camb} we compare the evolution of the dark matter perturbation calculated using CAMB in EDE models with wCDM. Normalising the growth factor to its present value, we see that at redshifts accessible to lensing tomography, $z<3$, the effect of typical EDE models on the growth of perturbations is at the percent level. However, at higher redshifts the presence of EDE is more apparent, and the integrated effect is significant. 

\subsection{The growth index parameterisation}

At late times, {\it e.g.}~$z<3$, the growth can be safely described by the growth equation  (\ref{eq:growth}).  The solution at these redshifts is well approximated using the well known growth rate formalism, or ``$\gamma$-index,''  and these approximate solutions can be a useful tool when considering late universe observations such as cosmic shear.  Specifically, in the standard $\Lambda$CDM scenario, $\Dplus(a)/a$ can be approximated very well through: 
\be
\Dplus(a)/a\ =\ \exp\left[ \int^{a}_{1} \frac{\Om(a')^\gamma-1}{a'} \dd a' \right]
\label{eq:g(a)}
\ee
where $\gamma=0.545$. When the dark energy is not a cosmological constant, this picture is modified somewhat.   A general dark energy equation of state parameter $w(a)$ will lead to a different expansion rate, and so to a different Hubble drag, leading to a shift in $\gamma$ \cite{Linder:2005in, Huterer:2006mva}.

More generally, $\gamma$ may be considered to be a function of the scale factor or other cosmological parameters. Wang \& Steinhardt \cite{Wang:1998gt} find
\be
\gamma(a)\ =\ \frac{3\left[1-w(a)\right]}{5-6w(a)}
\ee
to provide a good fit where $w(a)$ is the effective equation of state given by:
\be
w(a)\ =\ \frac{a\, \Om'(a)}{3\, \Om(a)\left[1-\Om(a)\right]} \,.
\ee
Here, $\Om(a)$ is
\be
\Om(a)\ =\ \frac{\Om^0}{\Om^0+\Od^0 a^{-3w_0}} \left[1-\Ode \left(1-a^{-3w_0}\right)^{2}\right] \,.
\ee
and the prime denotes the derivative with respect to the scale factor.  (While this expression for the matter density ignores the presence of radiation, it should be accurate for the low redshifts where this approximate solution is used.)  For a constant equation of state ({\it i.e.}~$\Ode=0$), the growth index defined in this way is independent of the matter density.   For more general dark energy models, and specifically when EDE is present, $\gamma$ weakly depends on the present matter density and the scale factor.  

\begin{figure}
\begin{center}
\includegraphics[width=0.65\textwidth]{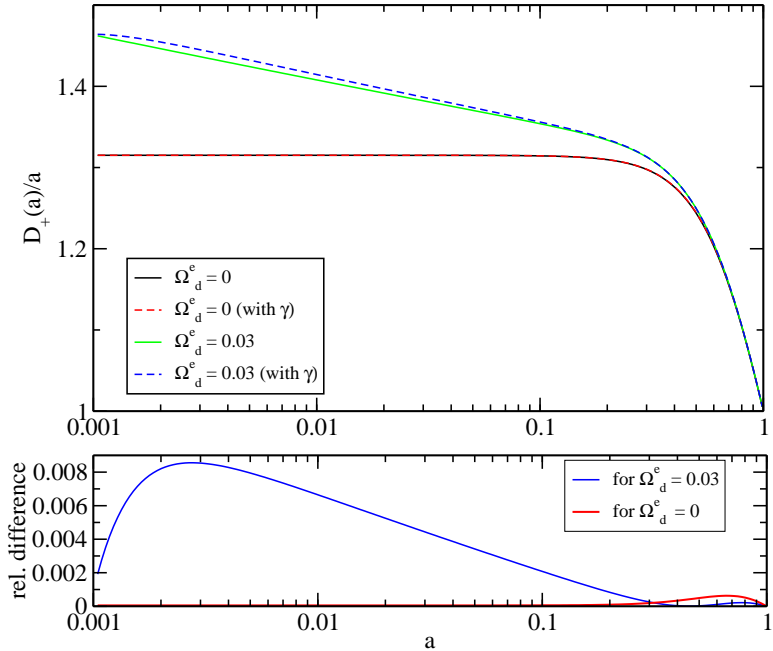}
\caption{In the top panel we plot the growth factor as a function of scale factor $a$ for wCDM and EDE with $\Ode=0.03$, comparing the numerical solution of the growth equation with the approximation using the growth index $\gamma$ given in equations (\ref{eq:growth}) and (\ref{eq:gamma}). In the bottom part we plot the relative differences and conclude that using the growth index is accurate at the $10^{-3}$ level.}
\label{fig:growth}
\end{center}
\end{figure}

Linder \cite{Linder:2009kq} has recently suggested an alternative approach for parameterising the approximate solution in EDE models, introducing an overall late-time calibration factor, $g_\star$, to account for the modified growth at early times due to EDE:
\bea
\gamma(w) &=& 0.55 + 0.05\left(1+w_{0}+\frac{5}{2}\Ode\right) \,,
\label{eq:gamma} \\
g_\star &=& 1-4.4\Ode \,.
\eea
We find this approach slightly more accurate in fitting the numerical solution of the growth equation. In \figref{fig:growth} we show the relative difference of the growth rate of dark matter perturbations obtained from the numerical solution of the growth equation and the approximation using \eqref{eq:growth} with $\gamma$ given (\ref{eq:gamma}). The accuracy is at the $10^{-3}$ level and therefore good enough for our purposes. 

\begin{figure}
\begin{center}
\includegraphics[width=0.65\textwidth]{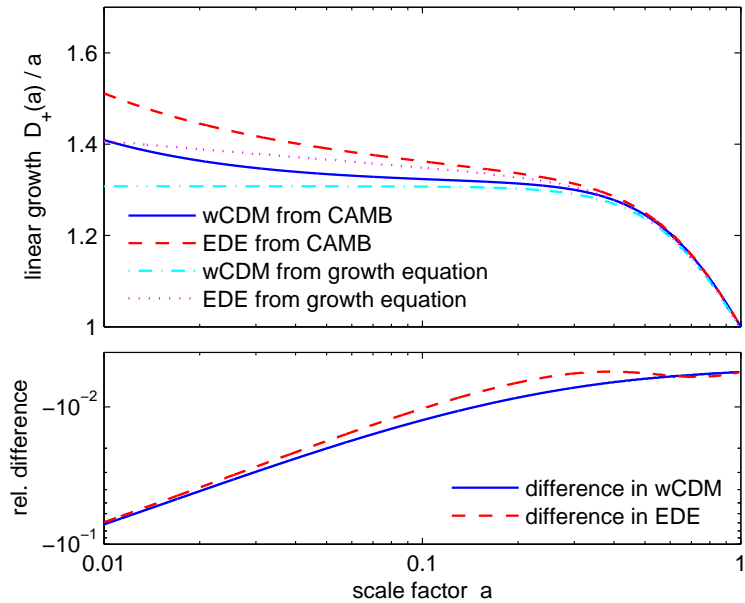}
\caption{Structure growth calculated by CAMB compared with the growth equation in a pure matter-dark energy universe (without radiation).  The error introduced by not including radiation is at the percent level for $a\lesssim 0.1-0.15$ (or $z\gtrsim 6-9$) depending on the cosmological model.}
\label{fig:rad_norad}
\end{center}
\end{figure}

Also shown in \figref{fig:rad_norad} is the growth as evaluated by the CAMB code, which takes into account the presence of radiation.  At late times the agreement is good, but the early growth is significantly altered.  Beyond a redshift of $6-9$, the radiation corrections become important at the percent level.

\subsection{Impact on the density power spectrum }

\begin{figure}[b]
\begin{center}
\includegraphics[width=0.495\textwidth]{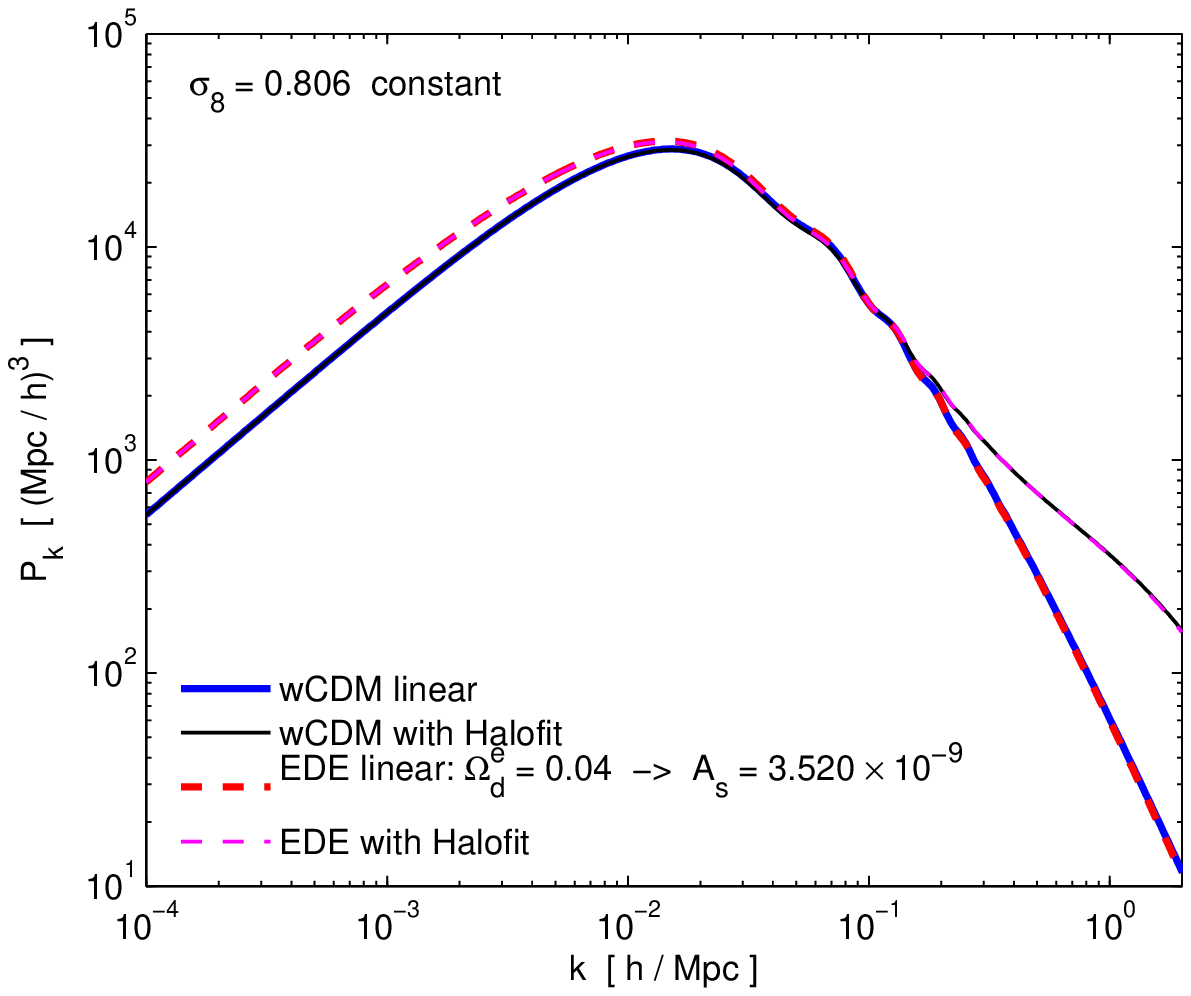}
\includegraphics[width=0.495\textwidth]{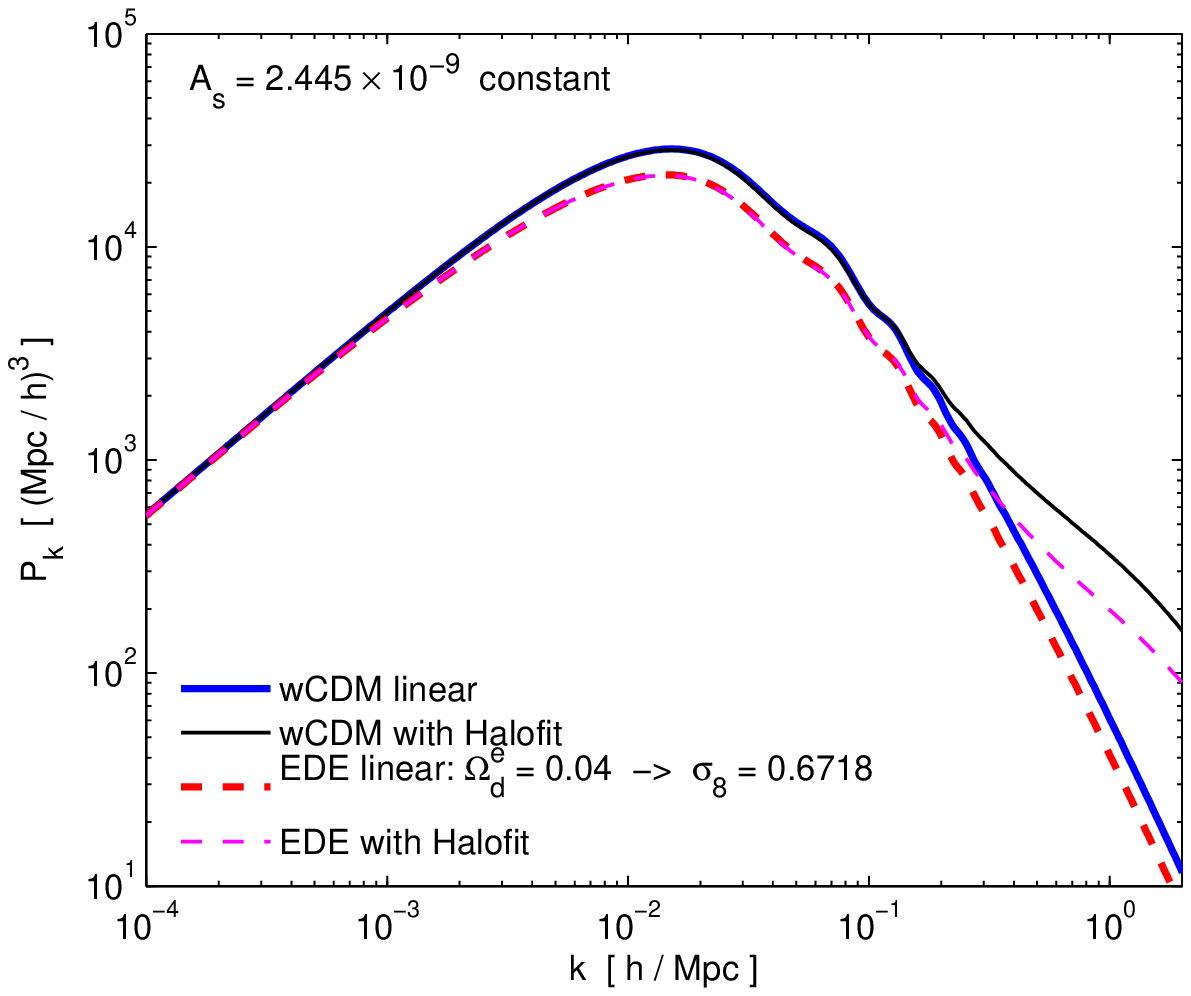}
\caption{The matter power spectrum is plotted for the WMAP+BAO+SN best fit wCDM model (blue, solid) and for the same model with EDE (red, dashed). We also plot the non-linear correction using the Halofit fitting formula based on N-body simulations by Smith \etal\cite{Smith:2002dz} for the best fit model (black, thin solid) and the EDE model (magenta, thin dashed). In the left panel $\sigma_8$ was fixed to the fiducial value by adjusting $A_s$ accordingly when EDE was added. In the right panel $A_s$ was held constant while $\sigma_8=0.6718$ for $\Ode=0.04$.}
\label{fig:ede_pk}
\end{center}
\end{figure}

The effect of EDE can be understood as a suppression of the growth of structure stronger than in $\Lambda$CDM which depends on the epoch when the scales enter the horizon.  For modes that enter the horizon before matter-radiation equality, $k>k_{\rm eq}$, all the modes experience the same suppression in growth.  For larger modes entering later, $k<k_{\rm eq}$, the suppression does not start until the mode enters the horizon which leads to a tilt in the power spectrum at large scales, see Caldwell \etal\cite{Caldwell:2003vp}. In \figref{fig:ede_pk} we plot the matter power spectrum for the fiducial model and compare it to the EDE model with the same cosmological parameters plus $\Ode=0.04$.

Although the shape of the matter power spectrum can be measured by galaxy surveys, the suppression due to EDE at the accessible scales is degenerate with the galaxy bias and the normalisation $\sigma_8$, as can be seen in \figref{fig:ede_pk}; thus, EDE cannot be constrained with these observations alone.  At the same time, nonzero $\Ode$ introduces a tilt in the power spectrum at large scales, $k<k_{\rm eq}$, and is therefore degenerate with the primordial spectral index $n_s$. Thus the combination of information on the growth of structure at different redshifts and at different scales is needed to determine a possible EDE component.

\section{CMB, CMB lensing and cosmic shear} \label{sec:theory}

\subsection{Information from the CMB}

The CMB is most sensitive to the physics of the early Universe, and observations of the CMB temperature and polarisation anisotropies are primarily useful for fixing the parameters, $n_s$, $\tau$, $\om$, $\ob$, as well as the primordial amplitude  $A_s$.  In the usual wCDM case, the sensitivity to the dark energy parameters ($w_0$ and $\Om$, which gives the present dark energy density) comes entirely through the observed angular scale of the sound horizon, since late dark energy changes the expansion and thus the distance to the last scattering surface. 

In EDE models, there is significant dark energy present even at last scattering.  The primary effect of this is to change the time of last scattering, and thus the actual sound horizon scale (in addition to affecting the distance to the last scattering surface) \cite{Doran:2000jt,Doran:2001yw}.  Since only one observable is affected, the angular scale of the sound horizon, the various dark energy parameters $\{\Ode$, $w_0$, $\Om\}$ are largely degenerate using CMB data alone.  (Note that the apparent sensitivity to $\Ode$ seen in the figures below is a result of our normalisation to $\sigma_8$, which is not probed directly by the CMB.) 
 
Information from CMB lensing and/or from other observational channels, like weak cosmic shear, are needed to break the degeneracies and measure these late Universe parameters accurately.   These observations provide a direct probe of the the present amplitude of perturbations, related to $\sigma_8$, and by combining these with the CMB constraints on $A_s$ one can constrain the amount of growth between last scattering and the present.  This provides a large lever arm to probe the suppression of growth caused by the presence of EDE.   Lensing measurements have the advantage that they are not subject to the bias issues which plague galaxy clustering observations.

\subsection{Information from CMB lensing}

CMB lensing is the gravitational deflection of CMB photons by the large scale structure between the last scattering surface and the observer. This secondary signal in the CMB has several statistical effects on the CMB anisotropies which are detectable. In this context we are interested in the changes to the power spectra and the generated $B$-mode polarisation, which can be used to extract cosmological information. See {\it e.g.}~Lewis \& Challinor \cite{Lewis:2006fu} for a comprehensive review of CMB lensing and Smith \etal\cite{Smith:2006nk} and Perotto \etal\cite{Perotto:2006rj} specifically on the estimation of cosmological parameters from the lensed CMB spectra.

The total deflection angle of a photon on its way to the observer is of the order of 2 arcminutes and therefore we might expect lensing to become an important effect at $l\gtrsim 3000$.  Unfortunately, on these angular scales, many foregrounds exist which could be problematic for the interpretation of the lensing signal.  However, the deflection angles are correlated over the sky by an angle of $\sim 2$ degrees, which is about the scale of the primary acoustic peaks in the CMB. This means that degree sized cold and hot spots appear after lensing larger or smaller by $\sim 2'$, or $\sim 3\%$. On average, lensing changes the statistics of the size distribution qualitatively, leading to a $\sim 3\%$ broadening of the acoustic peaks and dominates the spectrum on arcminute scales.

The lensing effect is described by re-mapping the CMB anisotropies by the deflection angle $\V{\alpha}$ such that the lensed temperature in direction $\V{\hat n}$, $\tilde{\Theta}(\V{\hat n})$, is given by the unlensed temperature in the deflected direction, $\tilde{\Theta}(\V{\hat n})=\Theta(\V{\hat n}+\V{\alpha})$. The tilde denotes lensed quantities. At linear order perturbation theory the deflection angle can be calculated as the gradient of a projected potential, $\V{\alpha}=\V{\nabla}\psi$. The so-called \emph{lensing potential} $\psi$ is a weighted integral of the gravitational potential along the line of sight and depends via the lensing efficiency function on cosmology.

\begin{figure}
\begin{center}
\includegraphics[width= 0.495\textwidth]{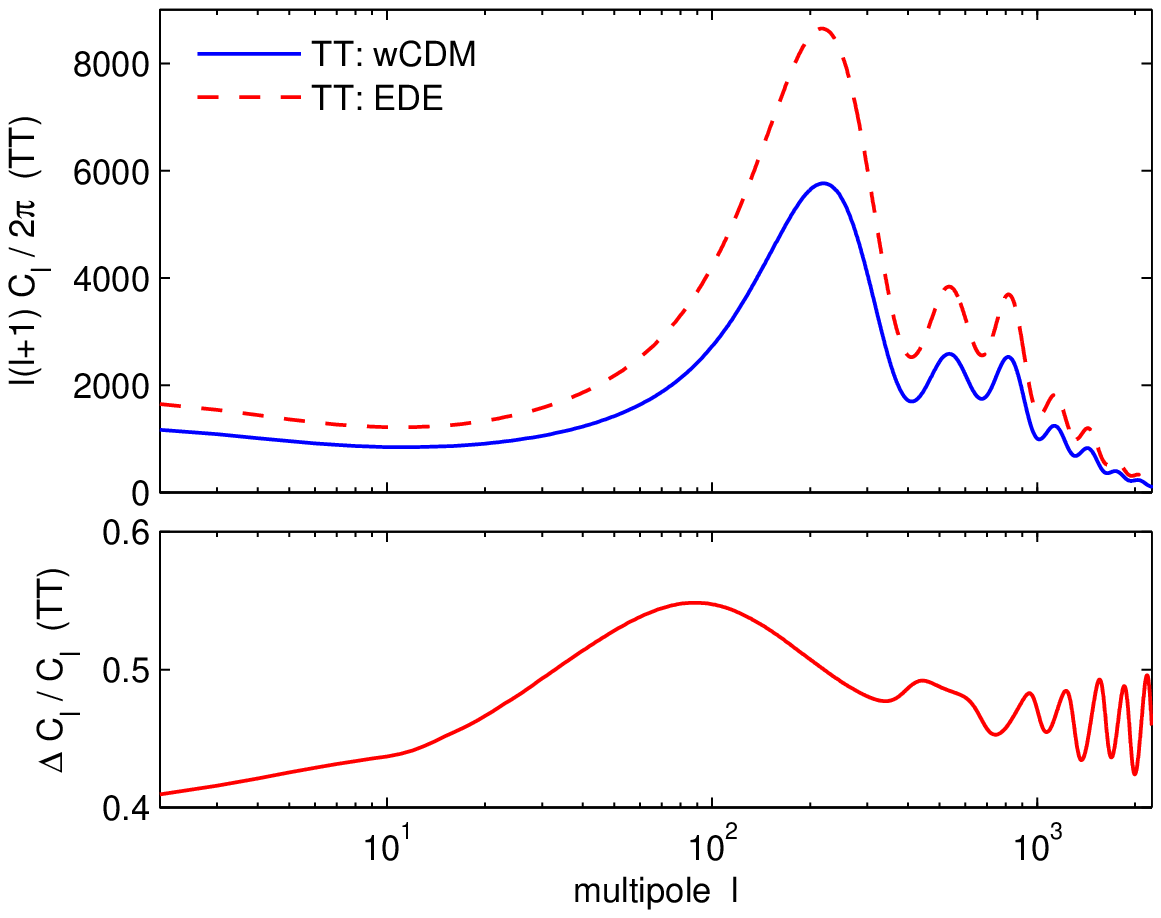}
\includegraphics[width= 0.495\textwidth]{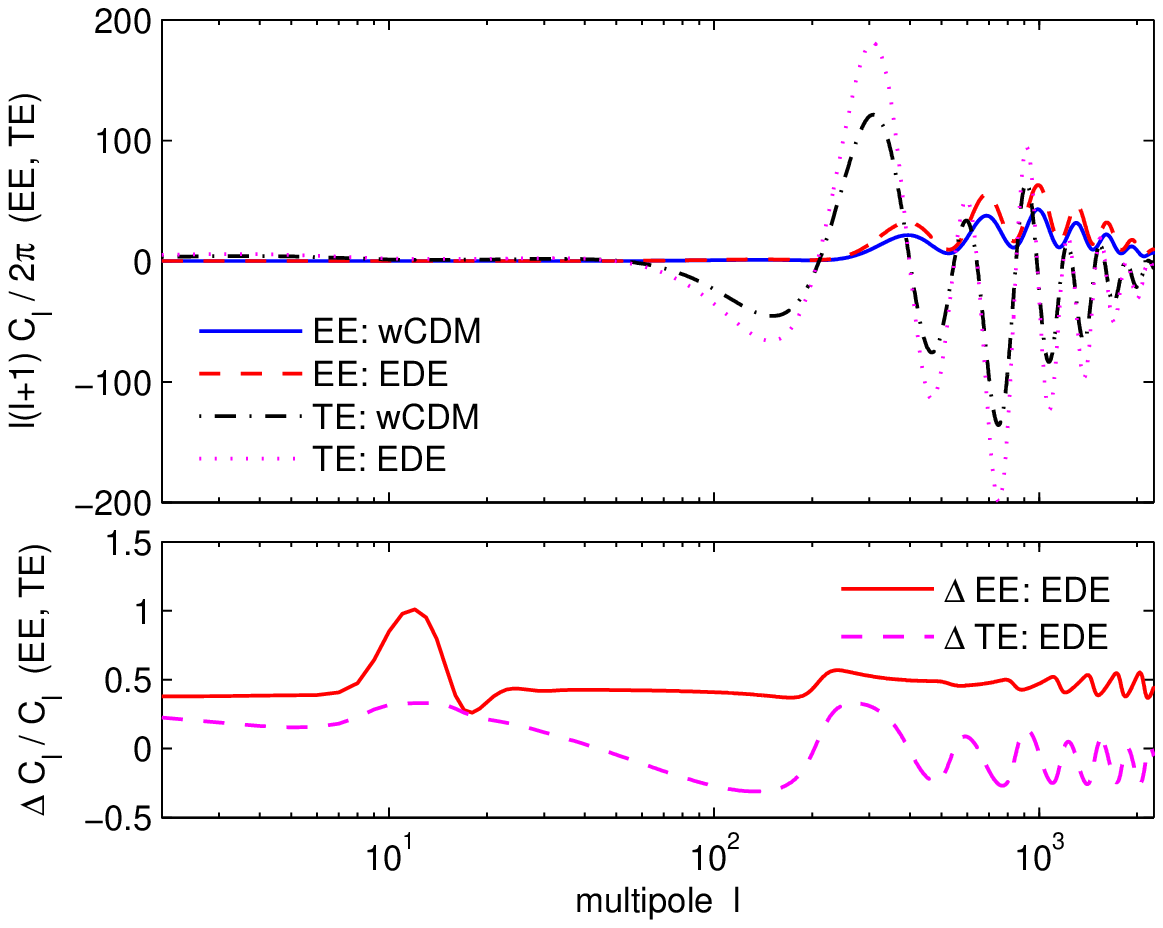} \\
\includegraphics[width= 0.495\textwidth]{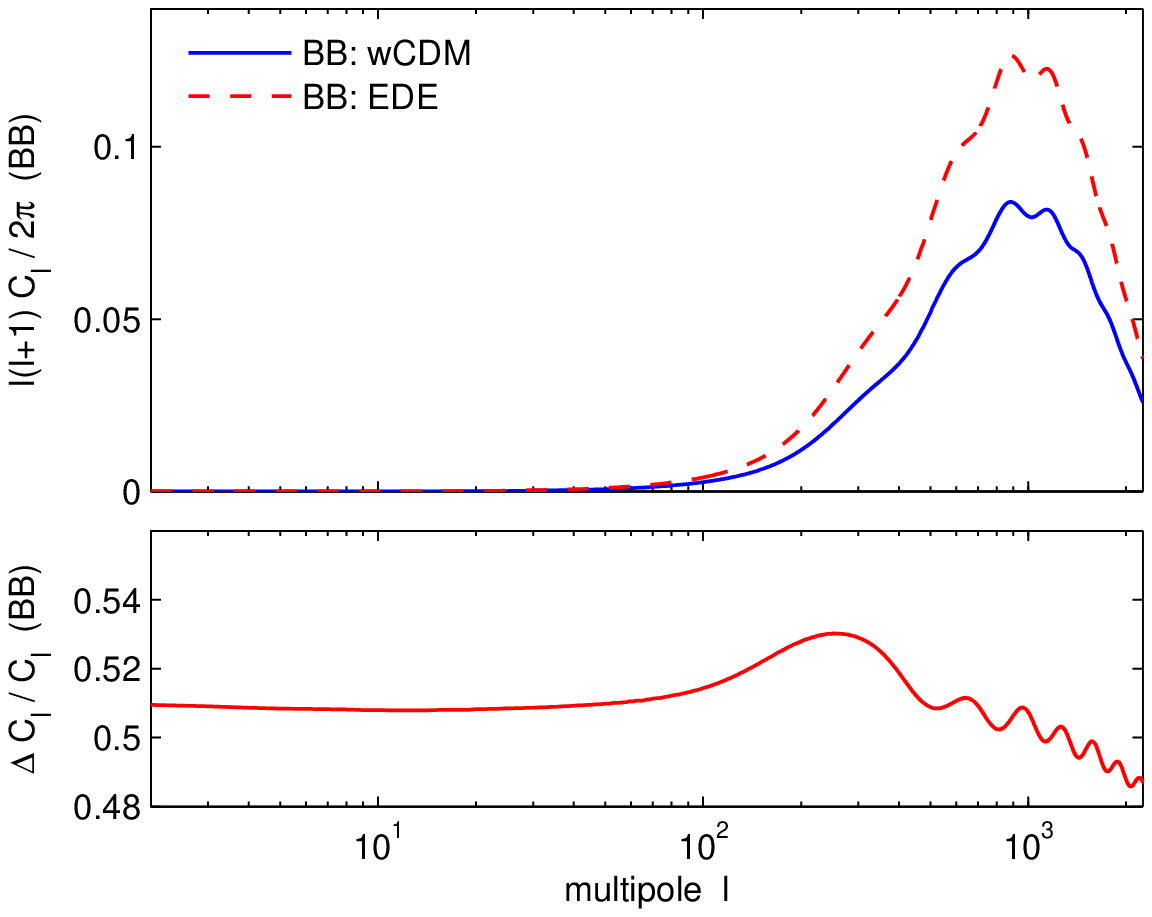}
\includegraphics[width= 0.495\textwidth]{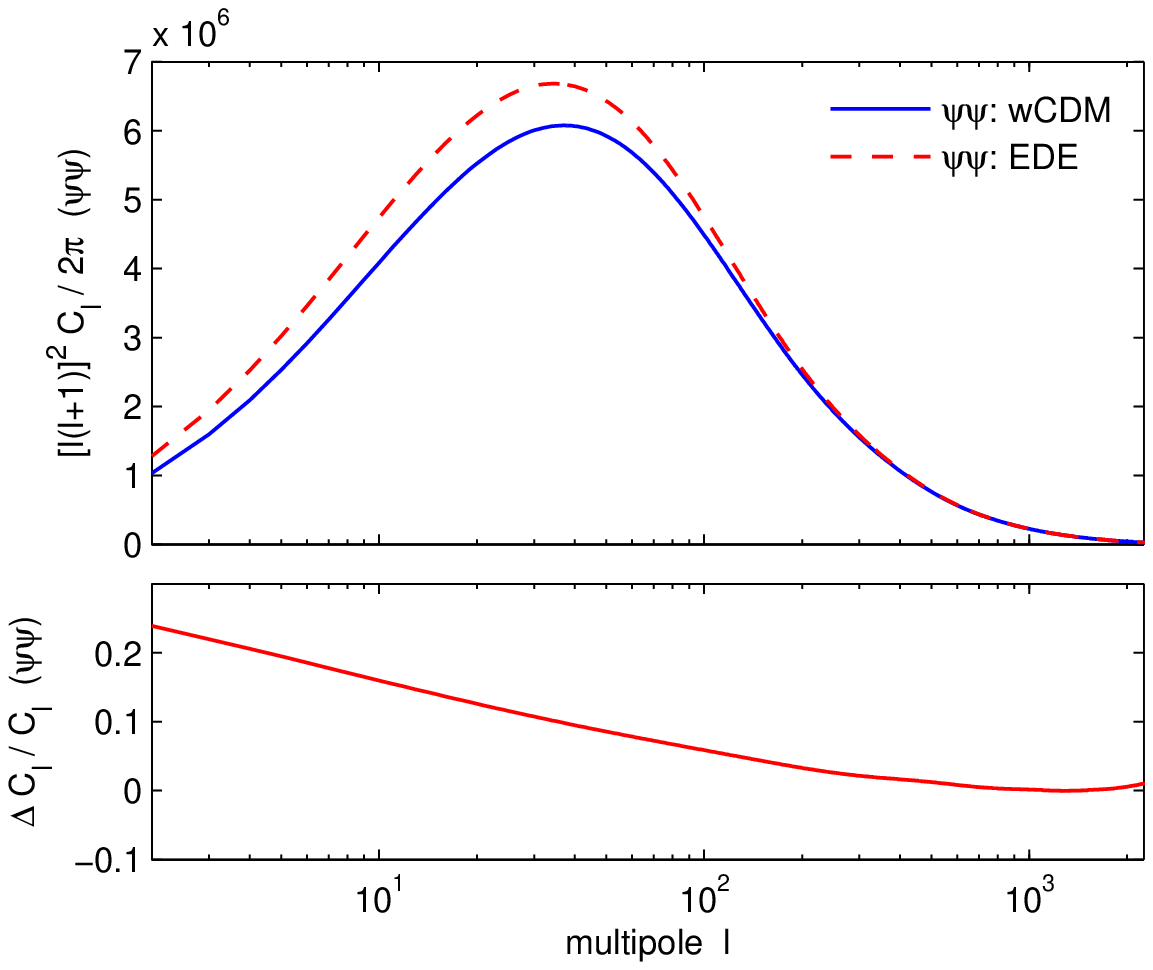}
\caption{The CMB power spectra for the WMAP+BAO+SN best fit wCDM model (blue, solid) are compared with those in case of EDE with $\Ode=0.04$ (red, dashed). Because we normalise to $\sigma_8$ today, the most apparent effect of EDE is an enhancement of power at the time of recombination.}
\label{fig:cls_ede}
\end{center}
\end{figure}

The interesting property of the lensing potential is that it has contributions from the structure out to quite high redshift. And in fact it is only weakly sensitive to late time and non-linear evolution. This characteristic makes CMB lensing a powerful tool for investigating EDE. In \figref{fig:cls_ede} we compare the primary CMB temperature and polarisation power spectra and the power spectrum of the lensing potential for the fiducial model with those in case we add a small amount of EDE. Because we normalise to $\sigma_8$ today, overall, EDE enhances the amplitudes of the CMB power spectra as the perturbations suffer a larger Hubble drag during their evolution. The lensing potential power spectrum is a weighted integral of the matter power spectrum and therefore also experiences a scale dependent enhancement at large scales (compare left panel of \figref{fig:ede_pk} and lower right panel of \figref{fig:cls_ede}).

The lensing potential, however, is not directly observable and thus, in an experiment, has to be estimated via reconstruction techniques using other observables. Lewis \& Challinor \cite{Lewis:2006fu} give an overview over the different possible techniques for the extraction of the lensing signal. All lensed auto- and cross-correlation power spectra of the temperature and the polarisation can be used to construct quadratic estimators for the lensing potential power spectrum as proposed by Zaldarriaga \& Seljak \cite{Zaldarriaga:1998te} and Guzik \etal\cite{Guzik:2000ju}. These are then combined to a \emph{minimum variance quadratic estimator} as introduced by Hu \& Okamoto \cite{Hu:2001kj} using a flat sky approximation. Okamoto \& Hu \cite{Okamoto:2003zw} construct it on the full sky which we implement to estimate the noise on the reconstructed lensing power spectrum.

\subsubsection{CMB Fisher matrix calculations}

To explore the power of CMB lensing to measure the growth history at intermediate redshifts and therefore to break the degeneracies between the late Universe parameters, we perform a Fisher matrix forecast for the upcoming Planck mission and later compare it with the futuristic proposal CMBPol. The Fisher matrix formalism is a straightforward and computationally fast method for predicting the precision with which future observations can measure cosmological parameters in a given model. The main assumption is that the parameter likelihood can be approximated by a Gaussian close to the maximum.

The Fisher matrix for a CMB experiment including lensing information is comprehensively presented by Perotto \etal\cite{Perotto:2006rj} and the results are compared to those resulting from sampling the likelihood by Monte Carlo Markov-chain methods. The errors derived from the Fisher formalism are in very good agreement with the Monte Carlo results for most model parameters. For strongly degenerate parameters and where parameter space is bounded from below or above the Fisher matrix is still found to yield a satisfactory estimate for an upper bound on the error.

The likelihood function ${\cal L}({\rm data}|\V{p})$ is approximated as a Gaussian function of the model parameters $\V{p}$. The likelihood is Taylor expanded about its peak at $\V{p}=\V{p}^0$. Then the Fisher information matrix is defined as the relevant second order term
\be
F_{\alpha\beta}\ \equiv\ -\left\langle\frac{\partial^2 \ln {\cal L}}{\partial p_\alpha\, \partial p_\beta}\right\rangle_{\V{p}^0} \,.
\ee
According to the Cramer-Rao minimum variance bound, the variance of the parameter $p_\alpha$, if marginalised over all other parameters, is given by:
\be
\sigma(p_\alpha)\ =\ \sqrt{\left(F^{-1}\right)_{\alpha\alpha}} \,.
\ee

For a CMB experiment the theoretical predictions are given in terms of the power spectra $C_\ell^{PP'}$ where $PP'\in\{TT,\ EE,\ TE,\ \psi\psi,\ \psi{}T\}$. Then the Fisher matrix can be expressed in terms of the derivatives of the theoretical power spectra with respect to the model parameters as
\be
F_{\alpha\beta}\ =\ \sum^{\ell_{\rm max}}_{\ell=2} \sum_{PP',\,QQ'}
\frac{\partial C_\ell^{PP'}}{\partial p_\alpha} \left(Cov_\ell^{-1}\right)_{PP'\,QQ'}
\frac{\partial C_\ell^{QQ'}}{\partial p_\beta} \,.
\ee
$Cov_\ell$ is the covariance matrix of the power spectra at multipole $\ell$. It is constructed from the observed power spectra, including all sources of variance to the fields, see Perotto \etal\cite{Perotto:2006rj} for details. For the CMB temperature and polarisation we add homogeneous instrumental noise and beam as given by Knox \cite{Knox:1995dq}
\be
N_\ell^{PP}\ =\ \left(\frac{\Delta_P}{T_{\rm CMB}}\right)^2 \exp\left[ \ell(\ell+1)\frac{\theta_{\rm FWHM}^2}{8\ln 2} \right]
\ee
where $\Delta_P$ is the detector noise for $P\in\{T,\ E,\ B\}$ and $\theta_{\rm FWHM}$ is the full width at half maximum of the beam. In the \tabref{tab:cmb_specs} we give the characteristics of the different CMB experiments considered. Generally we take the maximal multipole to be $\ell_{\rm max}=2250$ for both CMB experiments. Planck is noise dominated in all bands already for smaller multipoles and our errors from CMBPol will be an upper bound.

\begin{table}[hb!]
\begin{center}
\begin{tabular}[c]{| l | l | l |}
  \hline
    & Planck & CMBPol \\
  \hline
  temperature noise $\Delta_T$ &  28 \muka &  1 \muka  \\
  polarisation noise $\Delta_{E,\,B}$ &  57 \muka &  1.4 \muka \\
  beam $\theta_{\rm FWHM}$  &  7 arcmin &  3 arcmin \\
  fraction of sky $\fsky$  &  0.65  &  0.65 \\
  \hline
\end{tabular}
\caption{Assumptions on the characteristics of the different CMB experiments considered. Generally we take the maximal multipole to be $\ell_{\rm max}=2250$ for both CMB experiments.}
\label{tab:cmb_specs}
\end{center}
\end{table}

\begin{figure}
\begin{center}
\includegraphics[width= 0.495\textwidth]{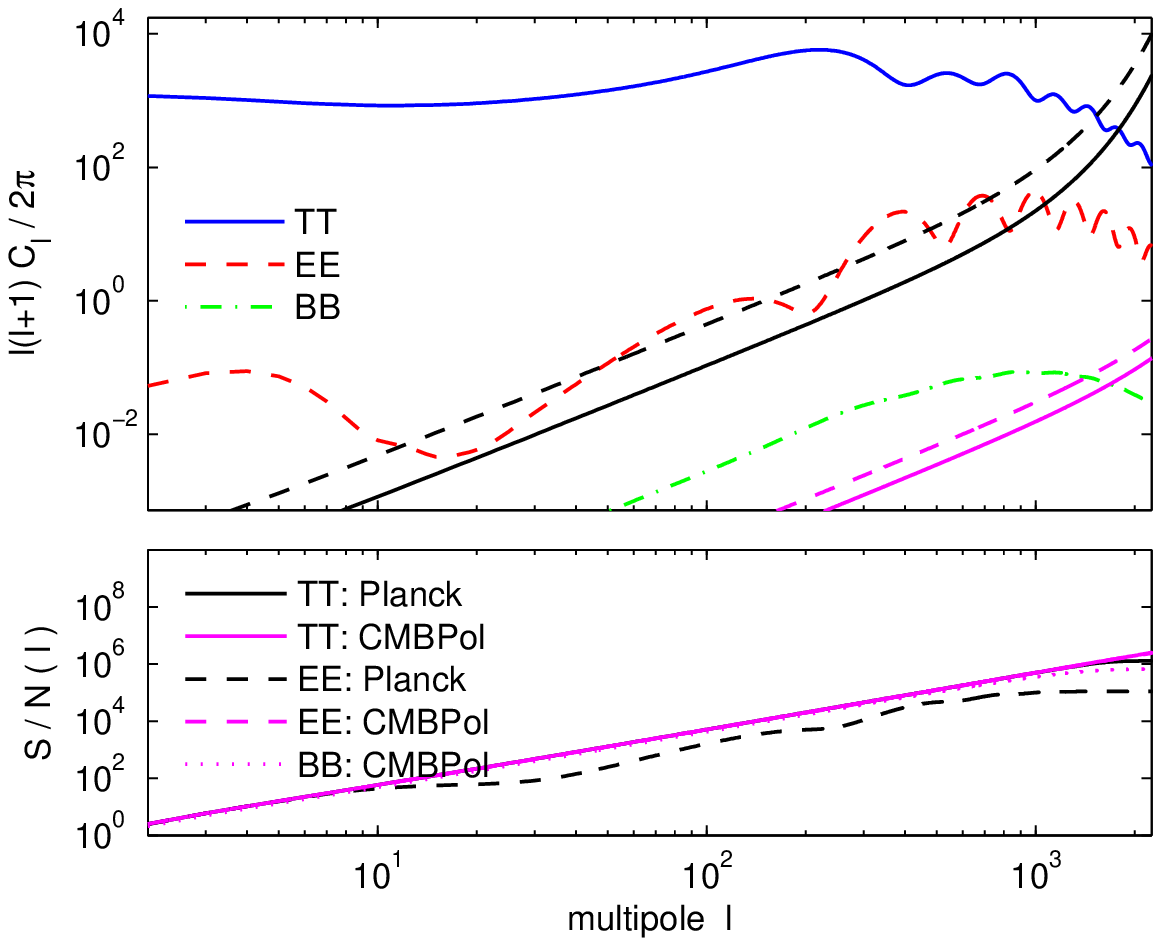}
\includegraphics[width= 0.495\textwidth]{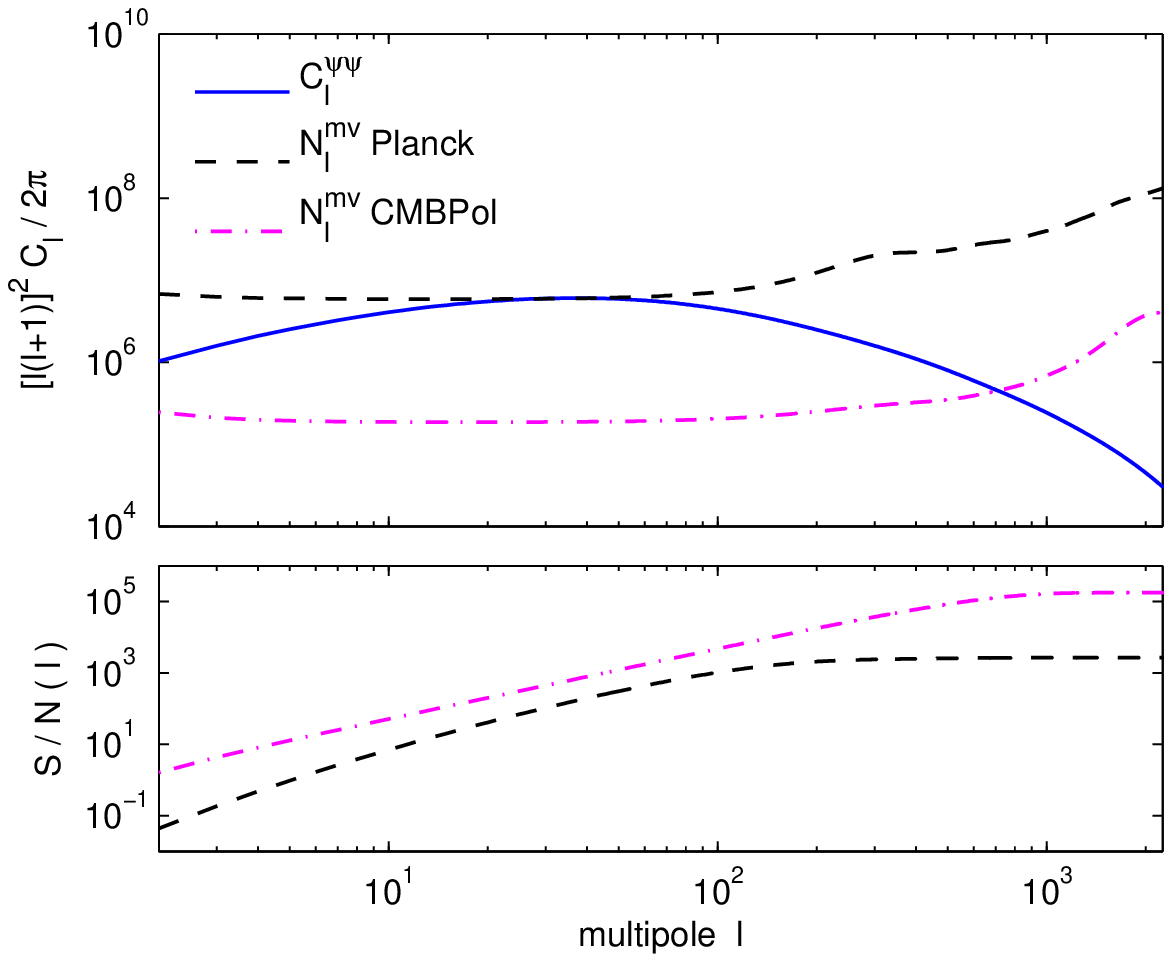}
\caption{The CMB power spectra for the WMAP+BAO+SN best fit wCDM model are plotted against the expected noise levels for Planck and CMBPol. The lower panels show the cumulated signal-to-noise as a function of the highest multipole.}
\label{fig:cls_noise}
\end{center}
\end{figure}

In \figref{fig:cls_noise} we plot the different theoretical power spectra for the fiducial model and compare them with the experimental noise. In case of the lensing potential power spectrum the noise is estimated as described above using the noisy temperature and polarisation power spectra given here. We also plot the cumulated signal-to-noise which we define as
\be
S/N(\ell)\ \equiv\ \sum_{\ell'=2}^{\ell} \frac{(2\ell'+1)}{2} \frac{C_{\ell'}}{C_{\ell'}+N_{\ell'}}\,.
\ee

To calculate the noise power spectra and the Fisher matrix for the Planck and CMBPol, we extended the Boltzmann code CAMB to include the EDE parameterisation defined in \eqref{eq:ede_param} and implemented the  lensing reconstruction on the full sky from Okamoto \& Hu \cite{Okamoto:2003zw}.

\begin{figure}
\begin{center}
\includegraphics[width=0.9\textwidth]{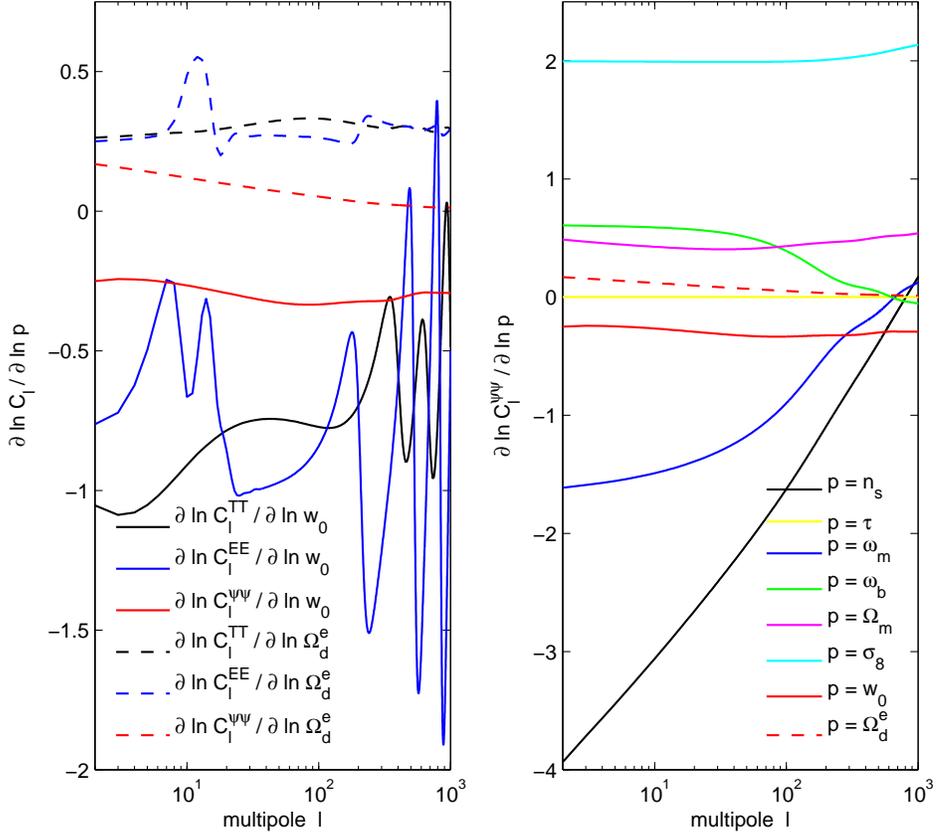}
\caption{\emph{Left panel:} logarithmic derivatives of the temperature, $E$-mode, and lensing potential power spectra with respect to $\Ode$ (dashed) compared to those with respect to $w_0$ (solid). \emph{Right panel:} logarithmic derivatives of the lensing potential power spectrum with respect to all model parameters.}
\label{fig:dCls}
\end{center}
\end{figure}

In \figref{fig:dCls} we plot, first, the derivatives of the primary temperature and $E$-mode power spectra with respect to the dark energy parameters $w_0$ and $\Ode$. Second, we plot the derivative of the power spectrum of the lensing potential with respect to all model parameters.

We checked the numerical stability of the derivatives (finite differencing) by varying the step size, comparing the usual double sided derivative with the one sided, and by using a 5-point derivative. No significant change in the results was observed. Furthermore, we calculated the results at different fiducial values for $\Ode$ and found this to be up to a $10\%$ effect on the marginalised errors.

\subsection{Information from cosmic shear}

We derive the sensitivity of typical next-generation tomographic weak lensing surveys to cosmological parameters following the recent literature \cite{Jain:2003tba, Ishak:2005we, Heavens:2006uk,Taylor:2006aw,Heavens:2007ka}. In particular, we study a Euclid type survey as laid out by Amara \& R\'efr\'egier \cite{Amara:2006kp}, which dealt with wCDM, and for modified gravity theories we refer to Amendola \etal\cite{Amendola:2007rr}.

Let us briefly recall the main equations for weak lensing studies. The weak lensing convergence power spectrum (which in the linear regime is identical to the ellipticity power spectrum) is a linear function of the matter power spectrum convolved with the lensing properties of space; and it can be written as \cite{Hu:2003pt}
\be
P_{ij}(\ell)\ =\ H_0^3\int_0^\infty \frac{H_0}{H(z)}W_i(z)W_j(z)P_{\rm nl}
\left[P_{\rm lin} \left(\frac{H_0\ell}{r(z)},\,z\right)\right] \dd z
\ee
where $P_{\rm nl}[P_{\rm lin}(k,z)]$ is the non-linear matter power spectrum at redshift $z$ obtained by correcting the linear power spectrum $P_{\rm lin}(k,z)$.  Here, the subscripts $i$ and $j$, refer to bins in redshift space of the lensed galaxies, which are most commonly separated using redshifts obtained photometrically (photo-$z$'s.) In flat space we have:
\bea
W_i(z) & = & \ \frac{3}{2}\Om F_i(z)(1+z)
 \,,\\
F_i(z) & = & \int_{Z_i} \frac{n_i(z_s) r(z,z_s)}{r(0,z_s)} \dd z_s
 \,,\\
n_i(z) & = & \ D_i(z)/\int_0^\infty D_i(z') \dd z'
 \,,\\
r(z,z_s) & = & \int_z^{z_s} \frac{H_0}{H(z')} \dd z'
\eea
where $D_i(z)$ is the radial distribution function of galaxies in the $i$-th redshift bin. We assume an overall radial distribution
\be
D(z)\ \propto\ z^2 \exp\left[-(z/z_0)^{3/2}\right] \,.
\ee
The distributions $D_i$ are obtained by binning the overall distribution and convolving with the photo-$z$ distribution function.

\subsubsection{Cosmic shear Fisher matrix calculations}

The Fisher matrix for weak lensing is given by
\be
F_{\alpha\beta}\ =\ \fsky \sum_{\ell}\frac{(2\ell+1)\Delta\ell}{2} \frac{\partial P_{ij}}{\partial p_\alpha} (C^{-1})_{jk} \frac{\partial P_{km}}{\partial p_\beta} (C^{-1})_{mi}
\ee
and the covariance matrix is defined as
\be
C_{jk}\ =\ P_{jk}+\delta_{jk}\left\langle \gamma_{\rm int}^{2}\right\rangle n_{j}^{-1}
\ee
where $\gamma_{\rm int}$ is the r.m.s.~intrinsic shear (we assume $\left\langle \gamma_{\rm int}^{2}\right\rangle ^{1/2}=0.22$ \cite{Amara:2006kp}) and
\be
n_{j}\ =\ 3600\, d\left(\frac{180}{\pi}\right)^2 \hat{n}_{j}
\ee
is the number of galaxies per steradian belonging to the $i$-th bin, $d$ being the number of galaxies per square arcminute and $\hat{n}_{i}$ the fraction of sources belonging to the $i$-th bin.

In the following we consider a satellite mission like Euclid with the characteristics given by R\'efr\'egier \cite{Refregier:2003ct}: sky fraction $\fsky=1/2$, mean redshift of the galaxy distribution $\zmean=0.9$, and number of sources per arcmin$^{2}$ of $d=40$. We assume that the photo-$z$ error obeys a normal distribution with variance $\sigma_{z}=0.05(1+z)$. We choose to bin the distribution out to $z=3$ into five equal-galaxy-number bins.

\begin{figure}
\begin{center}
\includegraphics[width=0.7\textwidth]{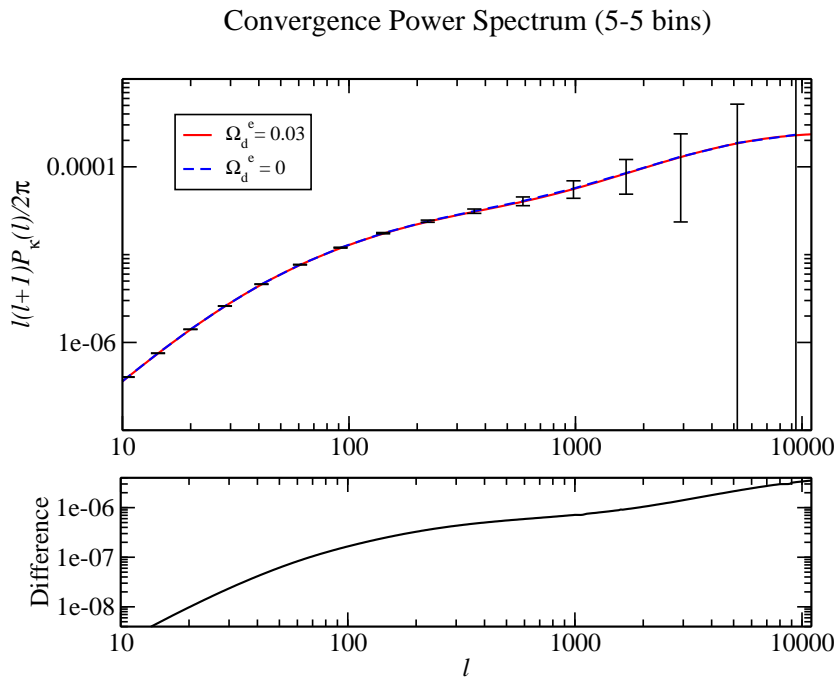}
\caption{The convergence power spectrum for the fifth redshift bin, $P_{55}(\ell)$, is shown for the WMAP+BAO+SN best fit wCDM model (blue, dashed) compared with the EDE fiducial model (red, solid). The lower panel shows the relative difference between the two. The error bars represent the noise errors on the EDE convergence power spectrum in logarithmically spaced bins. }\label{fig:Pij_55}
\end{center}
\end{figure}

\begin{figure}
\begin{center}
\includegraphics[width=0.7\textwidth]{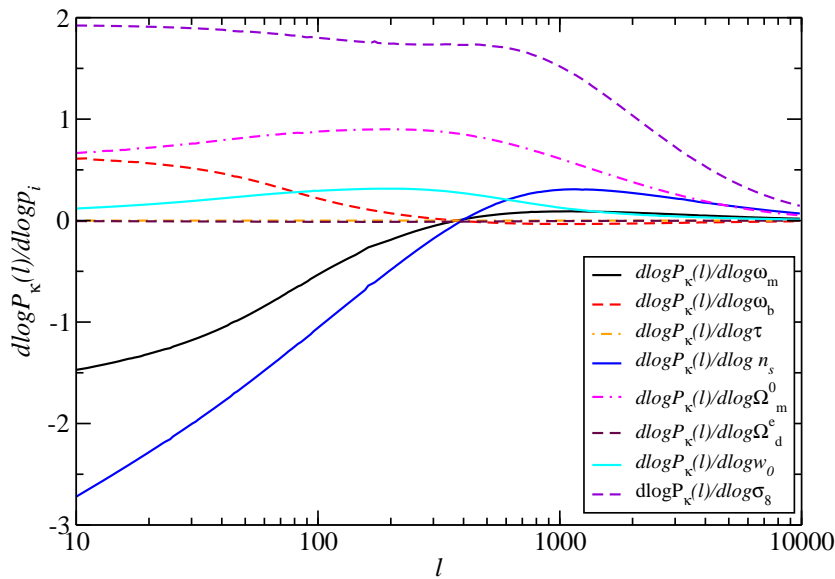}
\caption{Derivatives of the convergence power spectrum for the fifth redshift bin, $P_{55}(\ell)$ with respect to all model parameters.}
\label{fig:DerCnl55}
\end{center}
\end{figure}

In \figref{fig:Pij_55} we plot the convergence power spectrum for the fifth redshift bin, $P_{55}(\ell)$ for EDE fiducial model, with the noise due to the intrinsic ellipticity and for comparison we also plot the wCDM spectrum. We see that the EDE spectrum and the wCDM spectrum are almost undistinguishable. The derivatives of the same convergence power spectrum with respect to the model parameters are shown in \figref{fig:DerCnl55}.

For the linear matter power spectrum we adopt the fit by Eisenstein \& Hu \cite{Eisenstein:1997jh} (with no massive neutrinos and also neglecting any change of the shape of the spectrum for small deviations around $w=-1$). For the redshifts of interest, we verified that the Eisenstein \& Hu fit is accurate enough at the scales to which cosmic shear is sensitive. For the non-linear correction we use the fitting function to the halo model from Smith \etal\cite{Smith:2002dz}. We consider the range $10<\ell<10000$ since we find that both smaller and larger multipoles do not contribute significantly. 

For all results presented in the next section we assumed that the systematic errors can be neglected. A partial list of these errors include, on the observational side, shape measurement systematics and photometric redshift systematics and, on the theoretical side, the uncertainties concerning the accurate prediction of the small scale matter spectrum (due to non-linear effects and to the uncertain baryon contribution), the uncertainty on the source distribution, the degeneracies with other effects like massive neutrinos. Some of these effects have been shown to be under control, {\it e.g.}~the photometric redshift systematics and, to some extent, the shape measurements \cite{Amara:2006kp}.

\section{Results}  \label{sec:Results}

\subsection{Results for CMB and CMB lensing}

\begin{figure}
\begin{center}
\includegraphics[width=0.8\textwidth]{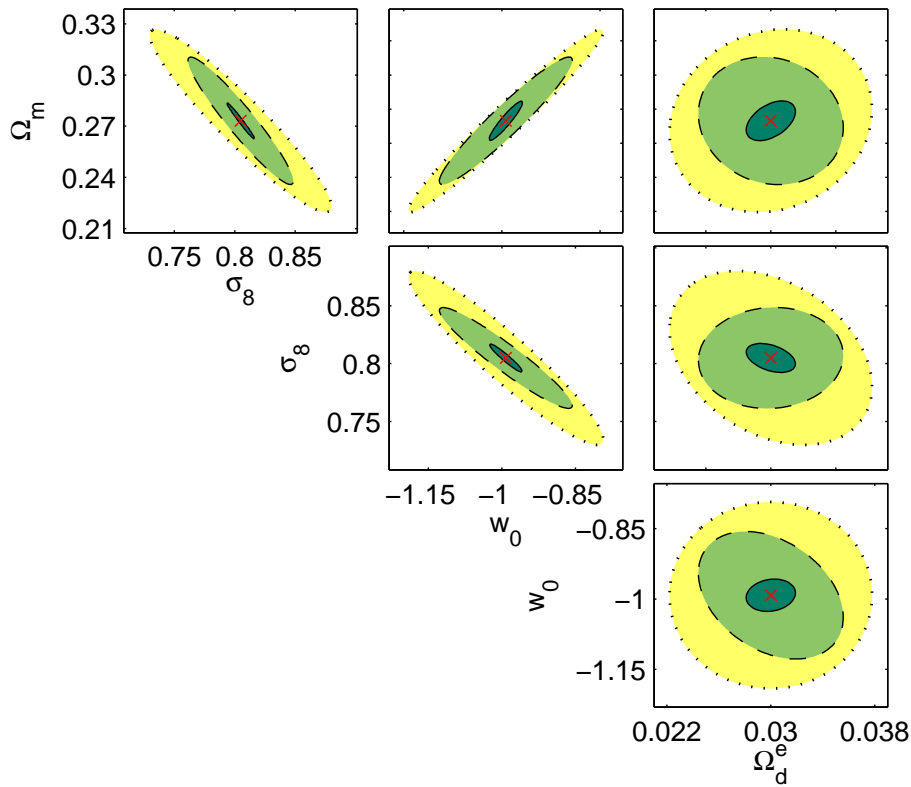}
\caption{Error ellipses (1$\sigma$) for all combinations between the late Universe parameters while marginalised over the other parameters. Projected constraints for Planck without CMB lensing are dotted (yellow), for Planck including CMB lensing are dashed (light green), and for CMBPol are solid (dark green) ellipses. Clearly already the CMB lensing information from Planck will help to break the degeneracies between late time parameters slightly.}
\label{fig:cmb_ellipses}
\end{center}
\end{figure}

In \figref{fig:cmb_ellipses} we present the forecasted $1\sigma$ error ellipses of all combinations of the model parameters for three cases: first, Planck without lensing information, second, Planck including the reconstructed lensing power spectrum, and last, CMBPol also including lensing information. The standard deviation on each single parameter (fully marginalising over all other parameters) is given in \tabref{tab:results} for the three cases.

The lensing information in Planck, as expected, does not improve the constraints on the early Universe parameters $\{n_s$, $\tau$, $\om$, $\ob\}$ significantly. On the other hand, it starts to break the degeneracy between the late Universe parameters $\{\Om$, $\sigma_8$, $w_0\}$ slightly. Finally, it improves the constraints on $\Ode$ by a factor $\sim 1.4$, and so Planck will be able to constrain EDE at sub-percent level, $\sigma(\Ode)=0.0037$.

CMBPol, the futuristic proposal of a dedicated satellite mission to map out the CMB polarisation modes, will obviously be able to constrain all parameters very well, as also pointed out by Smith \etal\cite{Smith:2008an} and de Putter \etal\cite{dePutter:2009kn}. The constraints on the single parameters are improved by factors $\sim 1.8-4$ compared to Planck with lensing.  Note that this result strongly depends on the achievable noise levels and the maximal multipole of CMBPol.

\subsection{Results for cosmic shear}

\begin{figure}
\begin{center}
\includegraphics[width=0.8\textwidth]{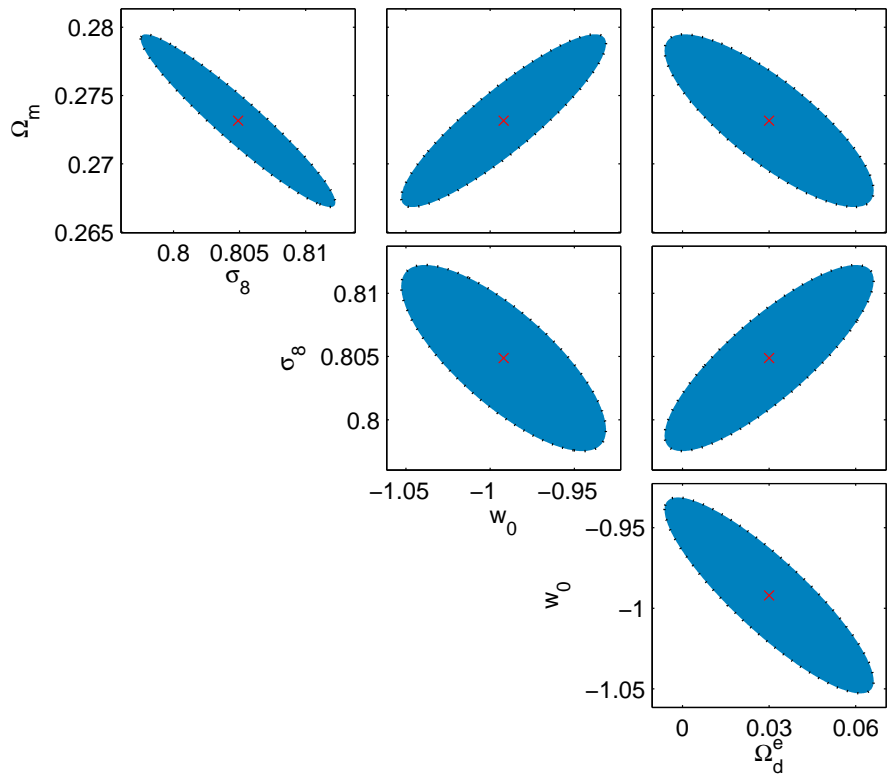}
\caption{Confidence regions at $68$\% for the a Euclid type lensing tomography survey with $z_{mean}=0.9$ and assuming 40 galaxies per arcmin$^2$.}
\label{fig:wl_ellipses}
\end{center}
\end{figure}

In \figref{fig:wl_ellipses} we show the confidence regions for the different combinations of the late Universe parameters $\{\Om$, $\sigma_8$, $w_0$, $\Ode\}$ . We always plot the fully marginalised confidence ellipses at $68$\%, which in two dimensions corresponds to semiaxes of length $1.51$ times the eigenvalue. Marginalising over all other parameters, we see that errors of $0.04$ for $w_0$ and $0.024$ for $\Ode$ are achievable with a Euclid type survey. In \tabref{tab:results} the $1\sigma$ errors on all model parameters are listed. Since the matter power spectrum does virtually not depend on the optical depth of reionisation a cosmic shear survey does not constrain $\tau$.

The predicted error $\sigma(\Ode)$ is comparable to the recent upper limit of $\Ode < 0.045$ by Bean \etal\cite{Bean:2002sm} indicating that galaxy weak lensing experiments are not strongly sensitive to an early dark energy component. On the other hand, the strong constraint on $w_0$ from cosmic shear tomography will help to constrain dark energy in combination with information from the CMB as we discuss in the next section.

In principle, one could consider a more nearer-term weak lensing survey, like CFHTLS Wide\footnote{The Canada-France-Hawaii Telescope Legacy Survey: \texttt{www.cfht.hawaii.edu/Science/CFHLS}} or DES\footnote{The Dark Energy Survey: \texttt{www.darkenergysurvey.org}}. However, their sky coverage is considerably smaller and therefore the statistical errors significantly larger. We verified that CFHTLS would not and DES would only marginally improve the constraints on EDE form Planck, respectively.

\subsection{Combined results}

We combine the Fisher matrices for the possible CMB, CMB lensing and cosmic shear measurements. The redshifts which are probed by CMB lensing and cosmic shear tomography are quite distinct from each other such that we can safely neglect the covariance between the measurements. This means we simply use the sum of the two Fisher matrices to calculate the combined constraints.

\begin{table}
\begin{center}
\begin{tabular}[]{| c || c | c | c | c || c || c | c |}
  \hline
  fiducial   & \multicolumn{7}{c |}{marginalised error $\sigma(p_\alpha)$} \\
  params.  &  Euclid & \multicolumn{2}{c |}{Planck} & \multicolumn{2}{c ||}{CMBPol}  &  Planck  &  CMBPol\\
                  & &  no lensing  &  lensing  &  no lensing  &  lensing  &  +Euclid  &  +Euclid
  \\ \hline \hline
  $n_s$      & 0.0094  & 0.0033  &  0.0031  & 0.0020   &  0.0018   & 0.0021   & 0.0013 \\
  $\tau$     &  -      & 0.0042  &  0.0040  & 0.0025   &  0.0019   & 0.0038   & 0.0019 \\
  $\om$      & 0.0059   & 0.0015  &  0.0015  & 0.00080  &  0.00064  & 0.00072   & 0.00032 \\
  $\ob$      & 0.0015  & 0.00013 &  0.00012 & 0.000042 &  0.000037 & 0.000093 & 0.000034 \\
  $\Om$      & 0.0042  & 0.035   &  0.025   & 0.0090   &  0.0077   & 0.0022   & 0.0020 \\
  $\sigma_8$ & 0.0049  & 0.050   &  0.029   & 0.013    &  0.0083   & 0.0024   & 0.0022 \\
  $w_0$      & 0.040    & 0.13    &  0.090   & 0.026    &  0.023    & 0.016    & 0.0080 \\
  $\Ode$     & 0.024   & 0.0051  &  0.0037  & 0.0025   &  0.0013   & 0.0022   & 0.00092 \\
  \hline
\end{tabular}
\caption{Forecasted errors (1$\sigma$) for the model parameters from the CMB: first, for a Euclid type space based cosmic shear survey, second, for Planck without lensing information, third, for Planck with lensing information reconstructed from the primary power spectra, fourth and fifth, for the futuristic proposal CMBPol with and without lensing reconstruction, then, for the combinations of Planck+Euclid and CMBPol+Euclid  (both incl.~CMB lensing). Note that Planck will already be able to constrain EDE to sub percent level and the combination with Euclid improves the constraints on $\Ode$ and $w_0$ to a level comparable with CMBPol.}
\label{tab:results}
\end{center}
\end{table}

\begin{figure}
\begin{center}
\includegraphics[width=\textwidth]{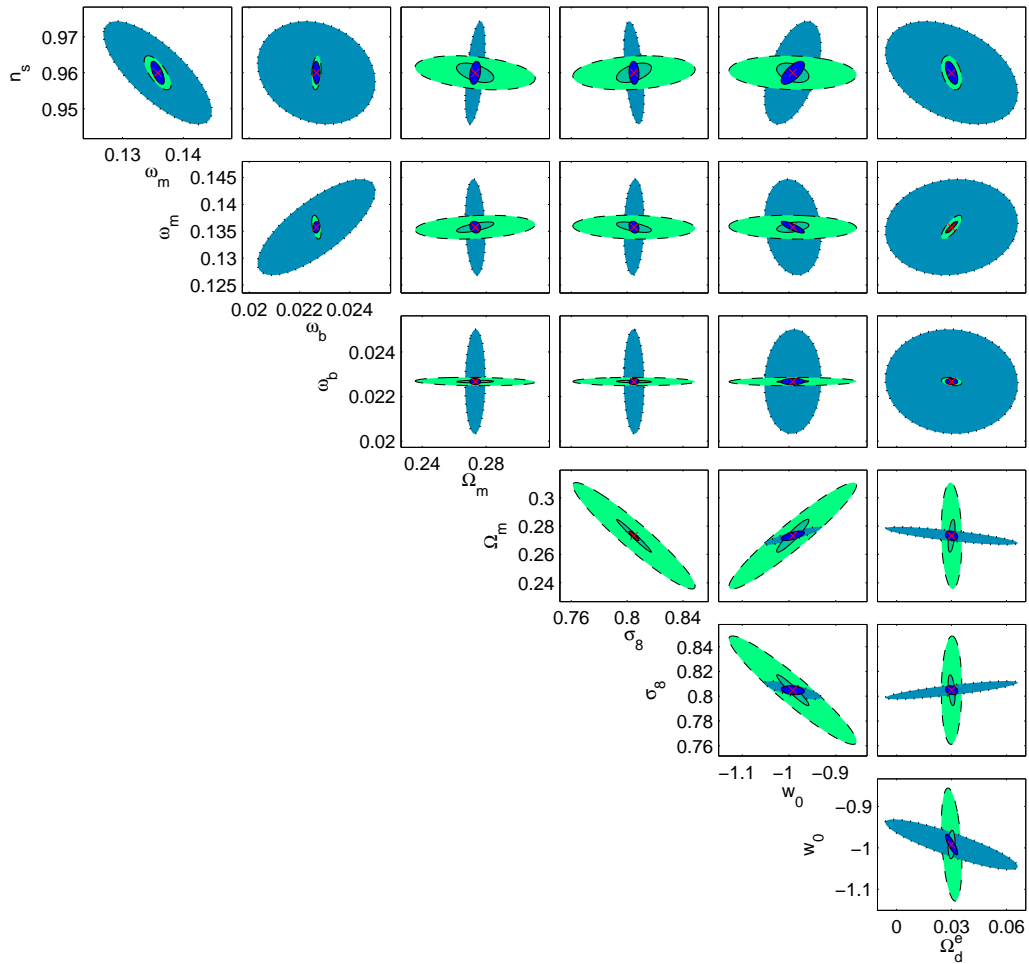}
\caption{Error ellipses (1$\sigma$) for all model parameters for CMB lensing from Planck (green, dashed), for cosmic shear from Euclid (turquoise, dotted), and for their combination (blue, solid).}
\label{fig:comined_ellipses}
\end{center}
\end{figure}

In \tabref{tab:results} we summarise the $1\sigma$ uncertainties derived for Euclid, Planck, and CMBPol, and for the combinations of Planck+Euclid and CMBPol+Euclid (both incl.~CMB lensing). The combinations clearly benefit from the information on the early Universe from the CMB and on the late Universe from cosmic shear. From Planck+Euclid we infer a constraint on EDE of $\sigma(\Ode)=0.0022$ when marginalised over all other parameters. This improves the constraint from Planck alone by a factor 1.7. In \figref{fig:comined_ellipses} we plot the $1\sigma$ confidence contours from Euclid, Planck, and Planck+Euclid for the different combinations of all parameters except $\tau$ which is not constrained by Euclid. 

\begin{figure}
\begin{center}
\includegraphics[width=0.65\textwidth]{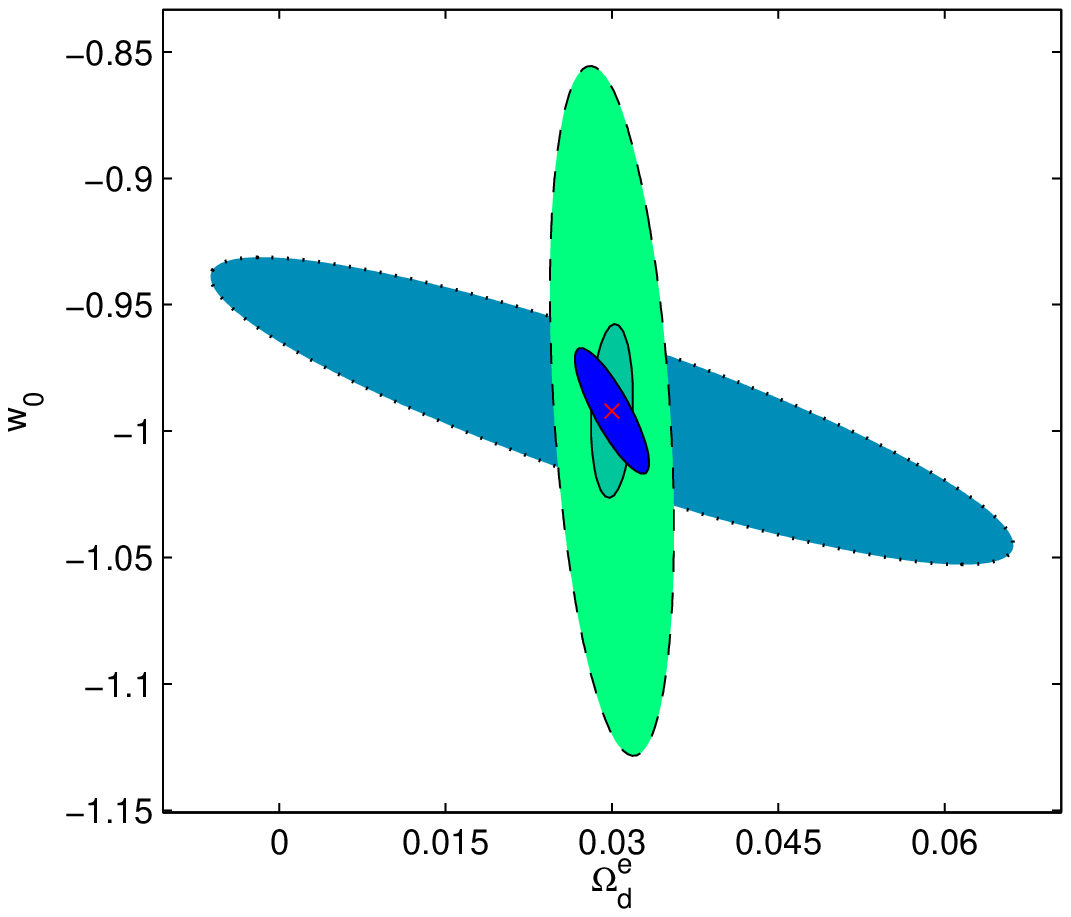}
\caption{Error ellipses (1$\sigma$) for the dark energy parameters $w_0$ and $\Ode$ from Planck (green, dashed), for cosmic shear from Euclid (dotted turquoise), for the combination Planck+Euclid (blue, solid), and from CMBPol alone (dark green, solid). The likelihood surfaces in the $(w_0,\Ode)$ plane for Planck and Euclid are strongly complementary which leads to a great enhancement of the constraints when combining the two probes.}
\label{fig:comined_ellipses_DE}
\end{center}
\end{figure}

In comparison, CMBPol alone will be able to constrain $\Ode$ down to $0.0013$. Overall, Planck+Euclid does quite well compared with CMBPol. This is, first, because the marginalised constraints on the late Universe parameters $\{\Om$, $\sigma_8$, $w_0\}$ from Euclid are comparable with those from CMBPol. And second, because the likelihood surfaces in the $(w_0,\Ode)$ plane for Planck and Euclid are strongly complementary as we point out in \figref{fig:comined_ellipses_DE}.  The combination CMBPol+Euclid would enhance all constraints even more.

\subsection{EDE figure of merit}

To be able to compare the leverage of the considered experiments on EDE we propose a figure of merit (FOM) based on the strength in constraining $w_0$ and $\Ode$ simultaneously while marginalising over all other parameters. We define $\tilde{F}^{-1}$ as the $2\times 2$ sub-matrix of the inverse Fisher matrix where the rows and columns associated with the marginalised parameters have been removed. Then we define
\be
FOM \equiv \sqrt{ \det \tilde{F} } \,.
\label{eq:def_fom}
\ee
In \tabref{tab:results_fom} the FOMs for the different experiments in consideration are listed together with the forecasted $1\sigma$ constraints on the dark energy parameters. Although CMBPol alone (incl.~lensing) does much better than Planck, the FOM of the combination Euclid+Planck is a factor $\sim 1.4$ higher than the FOM of CMBPol. This is because of the strong constraints from Euclid on the late-universe parameters which anchor the CMB and CMB lensing constraints at low redshifts.

\begin{table}[ht]
\begin{center}
\begin{tabular}[]{| c || c | c | c |}
  \hline
  Experiment   &  FoM  &  $\sigma(w_0)$  &  $\sigma(\Ode)$ \\
  \hline
  Euclid          &   2143   & 0.040   & 0.024 \\
  Planck          &   1500   & 0.13   & 0.0051 \\
  Planck lensing  &   3210   & 0.090  & 0.0037 \\
  CMBPol          &   16075  & 0.026  & 0.0025 \\
  CMBPol lensing  &   35629  & 0.023  & 0.0013 \\
  \hline
  Euclid+Planck   &   49765  & 0.016  & 0.0022 \\
  Euclid+CMBPol   &   157213 & 0.0080 & 0.00092 \\
  \hline
\end{tabular}
\caption{We compare the FOM defined in \eqref{eq:def_fom} and the $1\sigma$ errors on $w_0$ and $\Ode$ for the different experiments and the combinations of Euclid with Planck and CMBPol (both incl.~lensing). Note that Planck+Euclid has a higher FOM than CMBPol alone.}
\label{tab:results_fom}
\end{center}
\end{table}

\section{Summary \& conclusions}\label{sec:summary}

The focus of our investigation has been on early dark energy, its impact on the cosmic growth history and its observability with three structure formation probes, specifically the primary CMB temperature and polarisation spectrum at high redshift, the CMB lensing deflection field at intermediate redshift and weak cosmic shear at low redshift.

\begin{itemize}
\item{Early dark energy models show a scale dependent suppression of the cosmic density field during horizon entry and cause a tilt in the matter power spectrum, which is naturally degenerate with $n_s$ and $\sigma_8$. The Newtonian growth in early dark energy cosmologies is considerably higher than in standard dark energy models with varying equation of state. The growth function can be well approximated by the growth index.}

\item{The differences in the growth function at high redshift and the modifications of the power spectrum on large spatial scales suggest that the CMB fluctuations and CMB lensing are well suited for investigating early dark energy. Weak cosmic shear as a low redshift probe provides a precision measurement of $\Omega_m$, $\sigma_8$ and $w$ and is very useful for breaking degeneracies}

\item{We compute constraints on the early dark energy density and the degeneracy of $\Ode$ with other cosmological parameters from the CMB temperature and polarisation power spectra, with corresponding noise levels and angular resolution for the Planck surveyor and for the proposed CMBPol mission. The Fisher analysis suggests that the statistical uncertainty on $\Ode$ is $\sigma(\Ode)=0.0051$ from the CMB spectra observed by Planck, which decreases to $\sigma(\Ode)=0.0037$ including CMB lensing. The respective numbers for CMBPol are $\sigma(\Ode)=0.0025$ and $\sigma(\Ode) =0.0013$. We improve these measurements and break degeneracies by combining them with the Euclid weak lensing survey. By effectively adding priors on $\Omega_m$, $\sigma_8$ and $w_0$ at low redshifts, weak lensing data further decreases the uncertainty to $\sigma(\Ode)=0.0022$ for Planck and $\sigma(\Ode)=0.00092$ for CMBPol. The orientation of the degeneracy ellipses suggests that weak lensing is very well suited for breaking degeneracies and hence improving the constraints on $\Ode$ and $w_0$.}

\item{We define a figure of merit defined as the inverse volume of the parameter space spanned by $\Ode$ and the dark energy equation of state $w_0$ contained inside the $1\sigma$ isolikelihood contour. The increase of accuracy by combining CMB and lensing data is very illustrative: both Planck and Euclid for themselves reach values of about $2-3\times 10^3$, whereas their combination attains almost $5\times10^4$. A similar but lower value is reached for the CMBPol CMB polarisation power spectra, suggesting that a precision determination of these two cosmological parameters is possible with upcoming data sets.}

\item{In principle, other observational channels are also sensitive to $\Ode$. While the distance-redshift relation from supernovae is only expected to serve as a prior on the late Universe parameters, the integrated Sachs-Wolfe effect is mildly sensitive to the a possible early evolution of dark energy. However, this will be difficult to measure. More interesting will be future observations of the 21 cm radiation from reionisation with LOFAR and eventually with SKA. We expect a similar leverage on early dark energy from 21 cm lensing as from CMB lensing, see {\it e.g.}~Jarvis \cite{Jarvis:2007nw} and Metcalf \& White \cite{Metcalf:2008gq}.}
\end{itemize}

\section*{Acknowledgements}
D.S. acknowledges funding from the Swiss NSF and is grateful to Luca Amendola for interesting and helpful discussions. The work of B.M.S. was supported by the German Research Foundation (DFG) within the framework of the excellence initiative through the Heidelberg Graduate School of Fundamental Physics and by STFC during the early stages of this project.

\section*{References}
\bibliography{ede_refs}
\bibliographystyle{h-physrev4}

\end{document}